\documentclass[traditabstract]{aa}
\usepackage{txfonts}
\usepackage{epsfig}
\usepackage{graphicx}
\usepackage{booktabs}
\usepackage{natbib}
\usepackage{url}
\usepackage{twoopt}
\usepackage[breaklinks=true]{hyperref}
\begin{document}  

\title{Collisional effects in the  blue wing of the
  Balmer-$\alpha$ line\thanks{Opacity tables are only available at the CDS
    via anonymous ftp to cdsarc.u-strasbg.fr (130.79.128.5) } }

\author{
  N.~F.~Allard   \inst{1,2}\thanks{This paper is dedicated to the memory of Annie Spielfiedel and Roger
Cayrel for their fundamental work, rigor and kindness}
  \and
        F.~ Spiegelman\inst{3} 
  \and
  J. F. Kielkopf  \inst{4}
  \and
  S.~Bourdreux  \inst{2}
  }

   \institute{GEPI, Observatoire de Paris,  Universit\'e PSL, 
     UMR 8111, CNRS,
     61, Avenue de l’Observatoire, F-75014 Paris, France\\
          \email{nicole.allard@obspm.fr}\\
     \and
Sorbonne Universit\'e, CNRS, UMR7095, Institut d'Astrophysique
de Paris, 98bis Boulevard Arago, Paris, France\\
     \and
           Laboratoire de Physique et Chimie Quantique, F\'ed\'eration FERMI, Universit\'e de      
           Toulouse (UPS) and CNRS, 118 route de Narbonne, 
           F-31400 Toulouse,   France \\
\and
Department of Physics and Astronomy, 
    University of Louisville, Louisville, Kentucky 40292 USA \\
}

\date{Received 3june2021 / Accepted 20september2021; }

\abstract{
In order to investigate  the near wing of the
 Lyman-$\alpha$ line,    accurate line profile calculations and 
 molecular data  are both required  due to
 the existence of a close line satellite responsible for its  asymmetrical shape.   Lyman-$\alpha$ lines  observed 
 with the {\it Cosmic Origin Spectograph} ({\it COS}) on the {\it
   Hubble Space Telescope} ({\it HST}) show this peculiarity in the spectra of
 DBA and DA white dwarf stars.
  A similar asymmetrical shape in the blue wing can  be predicted in the  Balmer-$\alpha$ line of H perturbed by He and H atoms.

In continuation with a very recent work on the Lyman-$\alpha$ line, where the $n=2$ potential energies and 
transition dipole moments from the ground state were determined, 
we present    new accurate H-He potential energies and electronic transition dipole moments involving 
the molecular states correlated with H($n$=3)+He and their transition dipole moments with the 
states correlated with H($n$=2)+He. 
Those new data and existing molecular data for H($n$=2,3)-H are used  to provide a  theoretical investigation  of the
collisional effects in
  the blue wing of the  Balmer-$\alpha$ line of H perturbed by He and H atoms.
   We  note the consequences for the  Balmer-$\alpha$ line shape in 
  the physical conditions found in the cool atmosphere of
  DZA white dwarfs where
  helium densities may be as high as 10$^{21}$~cm$^{-3}$.
  This study is undertaken with  a unified theory of spectral line
broadening valid at very high helium densities.
}
 

\maketitle
\titlerunning{Collisional effects in  the  blue wing of Balmer-$\alpha$}
\authorrunning{Spiegelman, Allard, Kielkopf  \& Bourdreux}

\section{Introduction}
\label{sec:introduction}
Accurate atomic and molecular data, both theoretical and experimental,   are required to fully exploit current and future astrophysical space missions  and ground-based
facilities that deliver sensitive precise spectroscopy. Through the comparison of spectroscopic observations with models based on our best physics we gain an
understanding of the composition, temperature, density, magnetic fields, turbulence, and motions of the objects of interest.
An example, which  is the  key  point of this paper, is the decades-old
problem of the determination of the hydrogen abundance in
helium-dominated white
dwarf stars~\citep{cukanovaite2021}.
There is a discrepancy in the hydrogen
abundance determined from optical spectra of  Balmer-$\alpha$  and the abundance determined from ultraviolet spectra of
Lyman-$\alpha$ that is possibly
due to calculated opacities for these lines that have been incorporated into stellar atmosphere code.
 We have noted previously that there is a
 reliance on simplified models of neutral atom collision line-broadening physics, some dating to the work of \citet{unsold1955},
 in most of the  codes that are in
 current use. The reasons are understandable.
 Neutral collision broadening depends on  the knowledge of neutral atom
 interactions at separations typically greater
 than those at which atoms are stable in molecules.  At very large separations the van der Waals interactions that we can model apply and are
 well-understood, while   atoms in stars radiate in the presence of other atoms that are closer than this.  It is only relatively recently that the ab initio
   methodology for computing energies of atom pairs has developed the extreme precision needed to provide these data,
   especially for excited states. Consequently, potential energy models had been adopted
 that used ad hoc choices that were not validated in the laboratory. 
 For the determination of abundances, it is important that
 the  Lyman-$\alpha$  and Balmer-$\alpha$ lines be  fully resolved in
 low-noise observations, and
 be adequately represented in the theoretical models that are applied to them.
The use of  Lorentzian profiles with different widths predicted by the
hydrogenic van der Waals approximation  is inadequate to the task, as  was emphasized before by \citet{allard2007c} and \citet{peach2011}.
While an  accurate determination of the resonance broadening of Balmer-$\alpha$  has been  achieved by \citet{allard2008a},
nothing comparable exists for H-He.
The calculations reported in \citet{allard2008a} support the results
 of \citet{barklem2000b,barklem2002}
that the \citet{ali1966} theory of atom-atom resonance broadening for hydrogen underestimates the actual line width.
This will be the topic of a forthcoming paper; the present work is focussed on the blue wing of the Balmer-$\alpha$
and its asymmetrical shape due to radiation during close collisions.

In helium-dominated white dwarfs, the Lyman-$\alpha$ line profile is
asymmetric \citep[see][and references therein]{xu2017}.
The existence of a quasi-molecular line satellite  is crucial
      to   understanding  the asymmetrical shape of the
      Lyman-$\alpha$ line observed with the
      {\it Cosmic Origin Spectograph} ({\it COS}) on the {\it
    Hubble Space Telescope} ({\it HST})
      (see Fig.~1 in \citealt{allard2020}).
The resonance broadening of hydrogen perturbed by collisions with H atoms
produces asymmetry in the Lyman-$\alpha$  line profile similar to  that due
to H-He.
In a recent work, we investigated the absorption features in the blue wings
by computing detailed 
collisional broadening profiles \citep{spiegelman2021} for both H-He and H-H
using
new  Multi-Reference Configuration Interaction (MRCI) calculations of the  
 excited states potential energy curves of H-He dissociating into H($n$=2)+He, 
 as well as  the relevant electric dipole transition moments
 from the ground state
 contributing  to the Lyman-$\alpha$ spectrum.
 We  showed that
 tiny relativistic effects can affect the 
 asymptotic correlation of the H-He adiabatic states and change the related  transition dipole moments at  intermediate and long distance, significantly affecting the line profile.

 In this work we report  accurate potential energy curves  
 and transition dipole moments of H-He  involving initial and final states correlated with H($n$=2)+He and H($n$=3)+He, respectively.  
 Between these states, 16 H-He transitions  generate the complete Balmer-$\alpha$ 
 line profile. Although all of them were investigated for the present study,  
  we  restrict our discussion to the molecular data of
the  $\Sigma-\Sigma$ transitions which provide the essential contribution
to the blue wing (Sect.~\ref{sec:potHHe}),
and we  concentrate on the calculation of the
asymmetrical shape of the Balmer-$\alpha$ line perturbed by
He (Sect.~\ref{sec:BaHHe}) and H atoms (Sect.~\ref{sec:BaHH}).
In Sect.~\ref{sec:trend}  we  analyze
 the general trend of the repulsive $\Sigma$ excited states
which should lead to a non-Lorentzian shape  of the  Balmer lines when
we consider higher orders of the Balmer series in helium-rich white dwarfs.

\section{Diatomic H-He potentials and electronic transition dipole moments}
\label{sec:potHHe}

\begin{table}
\begin{center}
        \caption{\label{tab:hlevel}Comparison of calculated and experimental atomic energy levels of hydrogen (in cm$^{-1}$) for the
        $n$=2 and $n$=3 configurations} 
\vspace*{-0.3cm}        
\begin{tabular}{cccc}
\hline
        level  & Coulomb /DKH&Coulomb/DKH/mass & experimental\\
\hline
        &&&\\
        1s&0 &0&0\\
        2s &82303.923&82259.124&82258.954\\
        2p &82304.240&82259.440 & 82259.163\\
        3s &97545.453& 97492.357& 97 492.222\\
        3p &97545.607 & 97492.511& 97492.293\\
        3d & 97545.773&97492.678& 97492.341\\
\hline
\end{tabular}
\end{center}
{\footnotesize Column 2: Theoretical levels including the DKH
 contribution.  Column 3: Theoretical levels with the  DKH contribution
and finite proton mass correction. The experimental data 
in   Col. 4 correspond to   averages over the $j$
spin-orbit terms (weighted by $2j+1$) taken from  \citet{nist2020}.}
\end{table}

\begin{figure}
\centering
\vspace{8mm}
\includegraphics[width=8cm]{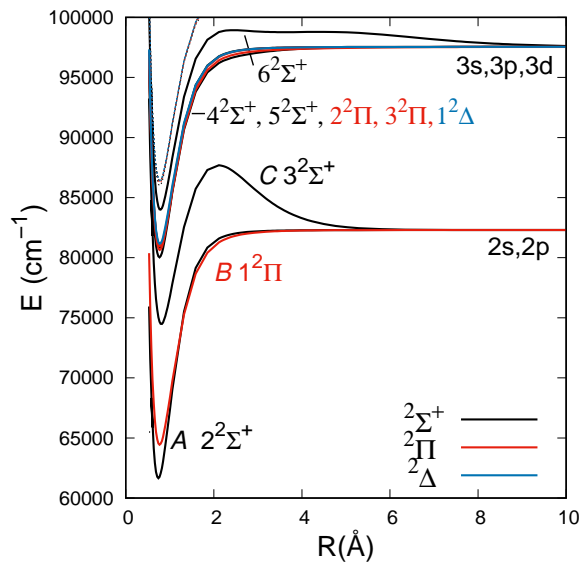}
        \caption{MRCI adiabatic potential energy curves of  HHe molecular states
  dissociating into  H($n$=2,3)+He.
  The thinner lines at the very top of the plot correspond to potential energy curves of bound states correlated with
  H($n$=4)+He, not dealt with   here.}
\label{fig:pothhe}
\end{figure}

\begin{figure}
\centering
\vspace{8mm}
\includegraphics[width=8cm]{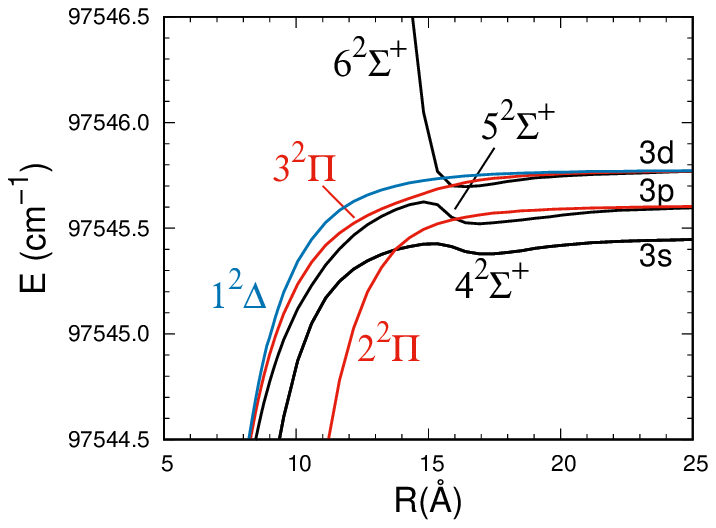}\\
\includegraphics[width=8cm]{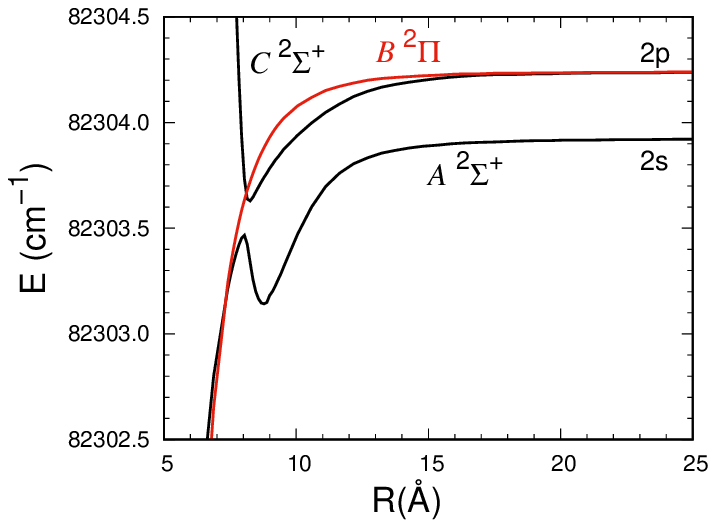}
\caption{Zoom-in on the long-range MRCI adiabatic potential energy curves
  of HHe. Top: States dissociating into  H($n$=3)+He.   
        Bottom: States dissociating into H($n$=2)+He. }
\label{fig:e3lr}
\end{figure}

\begin{figure}
\centering
\vspace{8mm}
\includegraphics[width=8cm]{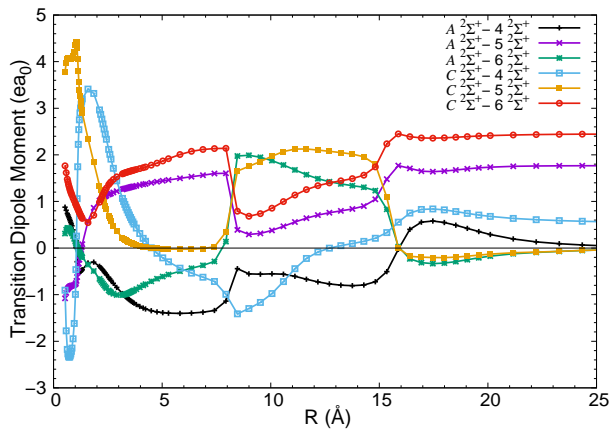}
\caption{Transition dipole moments of H-He between  
         $^2\Sigma^+$ states $n$=2 and  $n$=3.}
\label{fig:dipsig23}
\end{figure}

The calculation scheme for   the potentials dissociating
into H($n$=2,3,4)+He was described in 
our recent publication dealing  with  the Lyman-$\alpha$ line
broadening \citep{spiegelman2021}. Briefly, a MRCI calculation
\citep{knowles92,molpro2015} was run within an extensive Gaussian-type
orbital (GTO) basis set, namely  239 GTOs  on He and 297 GTOs on H. 
 Scalar relativistic corrections, including the Darwin and the
  mass-velocity contributions in the Douglas-Kroll-Hess scheme (DKH) at
  second order \citep{reiher2006,nakajima2011} were added, breaking the
  specific degeneracy of the atomic hydrogen levels determined with the
  Coulomb Hamiltonian only. 
 We showed in our previous work
\citep{spiegelman2021} that although their influence on the energy was less
than 1~cm$^{-1}$,  accounting for  relativistic effects could induce
long-distance avoided crossings, in turn resulting in  changes in the asymptotic
correlation of the adiabatic states and  in the dipole transition moments.
 Table~\ref{tab:hlevel} shows that the basis set used for 
  the $n=2$ and $n=3$ describes the energy levels of hydrogen with an accuracy better than
  0.5~cm$^{-1}$ compared to   the $j$-averaged experimental 
levels  deduced  from \citet{nist2020}.
In the case of the  H-He system, all energy eigenvalues and eigenstates
were obtained up to H($n$=4)+He (the $n$=4 manifold is not discussed here). 

In the following we  use the  spectroscopist's notations $X$, $A$, $B$, and $C$
for the lower adiabatic molecular states of H-He,  namely the ground state $1~^2\Sigma^+$ and  the lowest excited states $2~^2\Sigma^+$,
$1~^2\Pi$, and $3~^2\Sigma^+$ correlated with H($n$=2)+He, while we will
label  the upper states correlated with H($n$=3)+He according to their 
adiabatic ranking in their  respective symmetries. 
 In the literature, states $4~^2\Sigma^+$, $2~^2\Pi$,
and $5~^2\Sigma^+$ are also labeled $D$, $E$, and $F$, respectively.
Figure~\ref{fig:pothhe} shows the potential energy curves of the states
dissociating into H($n$=2)+He and H($n$=3)+He.
The lowest excited states ($n$=2) were abundantly discussed
 by \citet{theodora1987},
\citet{ketterle85}, \citet{ketterle88}, \citet{brooks88}, \citet{ketterle89}, \citet{vanhemert91} , \citet{petsalakis92}, \citet{sarpal1991}, \citet{lo2006}, \citet{allard2020} and \citet{spiegelman2021}.
 
  The three states $A$, $B$, and $C$ are bound at short distance
  ($R_e\approx$ 0.75~\AA\/ for states $A$ and $B$  with respective
  dissociation energies 20758 and 17868~cm$^{-1}$),
  however, with a smaller dissociation energy ($D_e$=13665~cm$^{-1}$) and a
slightly larger equilibrium distance ($R_e$=0.8093~\AA\/) for state
$C~^2\Sigma^+$, correlatively with a large potential energy  barrier to
dissociation (data taken  from \citealt{spiegelman2021}).
States dissociating into H($n$=3)+He have a similar behavior;
specifically,  all states are  bound with essentially parallel potential energy curves
having  equilibrium distances  $R_e$ around 0.77~\AA\/ and dissociation
energies $D_e$ in the range 16300--17600~cm$^{-1}$. However, the highest state
6~$^2\Sigma^+$ is  peculiar. It has  a  long-distance  barrier and
is significantly less bound ($D_e$=13554~cm$^{-1}$) than the former states and
at slightly larger distance $R_e$=0.7851~\AA\/. Starting from dissociation,
the barrier becomes significant below 10~\AA\/ and extends over a large range
down to 1.5~\AA\/. 
It presents a double hump structure, the inner maximum being the highest one
(1390~cm$^{-1}$ above the asymptote around $R$=2.1~\AA\/), significantly lower
than that of the $C~^2\Sigma^+$ state correlated with He(2$p$)+He
(5386~cm$^{-1}$ above the asymptote around $R$=1.90~\AA\/). 
Detailed adiabatic correlation of the molecular states of H-He towards
the atomic states is illustrated in 
the magnified image in Fig.~\ref{fig:e3lr} (top panel) showing the  long-distance
 behavior of the potential curves. In the range 14-16~\AA\/, states
4~$^2\Sigma^+$, 5~$^2\Sigma^+$, and 6~$^2\Sigma^+$  undergo sequential
avoided crossings and also crossings with states 2~$^2\Pi$,  3~$^2\Pi$, and
1~$^2\Delta$. 
Although more states are involved in the $n$=3 manifold, the situation is
similar to that of the H($n$=2)+He states shown in
Fig.~\ref{fig:e3lr} (bottom panel)   where the crossings occur at around
8.1~\AA\/. 
 The spectroscopic constants of the  adiabatic
states are
given in  the Appendix and are  compared with the literature values
\citep{theodora1987,sarpal1991,ketterle85,ketterle88,ketterle89,ketterle90a,ketterle90b,ketterle90d,vanhemert91,lo2006}.
The $rms$ deviation of the calculated $v'=0$ to $v"=0$ transition energies
($T_{00}$ in the Appendix table) from the experimental values of \citet{ketterle90b} is 23~cm$^{-1}$. No asymptotic shift or any empirical correction has been made for their determination. However, the inter-state non-adiabatic perturbations \citep{vanhemert91,petsalakis92} are not accounted for in the present work. 

All 16  transition dipole moments between the $n$=2 and $n$=3 adiabatic
states were calculated, namely transitions from $A~^2\Sigma^+$,
 $C~^2\Sigma^+$ to 4,5,6~$^2\Sigma^+$ and to 2,3~$^2\Pi$,  and those  from
 $B~^2\Pi$ to 4,5,6~$^2\Sigma^+$, 2,3~$^2\Pi$ and  1~$^2\Delta$.
In Fig.~\ref{fig:dipsig23} we focus on  the \mbox{$^2\Sigma^+$-$^2\Sigma^+$} transition dipole
moments since mainly the \mbox{$C~^2\Sigma^+-6~^2\Sigma^+$}
transition is  involved in the blue wing contribution, as discussed below.
Transition moments start to depart smoothly from their asymptotic values below
25~\AA\/, as  can be seen in
Fig.~\ref{fig:dipsig23}. Their evolution with distance can be understood
considering features characterizing  either the upper or the lower states  of  the
transitions. The first feature is  the avoided crossings between
4~$^2\Sigma^+$, 5~$^2\Sigma^+$, and 6~$^2\Sigma^+$  around R=16~\AA\/, which
causes switches betwen
$A~^2\Sigma^+-4~^2\Sigma^+$, $A~^2\Sigma^+-5~^2\Sigma^+$, and $A~^2\Sigma^+-6~^2\Sigma^+$
dipole moments on the one hand, and between  $C~^2\Sigma^+-4~^2\Sigma^+$,
$C~^2\Sigma^+-5~^2\Sigma^+$ and $C~^2\Sigma^+-6~^2\Sigma^+$ dipole moments on the
other hand. The second feature is  the avoided  
crossing between states $A~^2\Sigma^+$ and   $C~^2\Sigma^+$ around R=8.1~\AA\/,
which causes sharp switches between $A~^2\Sigma^+-4~^2\Sigma^+$ and
$C~^2\Sigma^+-4~^2\Sigma^+$,
$A~^2\Sigma^+-5~^2\Sigma^+$ and C~$^2\Sigma^+-5~^2\Sigma^+$, and
$A~^2\Sigma^+-6~^2\Sigma^+$ and $C~^2\Sigma^+-6~^2\Sigma^+$, respectively.
Finally, we can also observe avoided crossings of states  in their repulsive
inner branch  for R $<$ 0.7~\AA\/ that induce abrupt variations
in the transition moments.

While relativistic effects, partly taken into account here, obviously play a negligible role on the energies
(less than 1 cm$^{-1}$) and in the transition dipole moment
at short distance, they break the asymptotic degeneracy  and  induce
cascade avoided crossings
which govern the asymptotic correlation of the adiabatic states and the
variation of the dipole transition moments at medium and long distance.
For a full spectral account
of the  perturbation in the line profiles, it would have been interesting
to include  spin-orbit coupling (spin-orbit coupling  is of the
same magnitude as the scalar contributions) at the expense of dealing with more molecular transitions,
still more complex features in molecular data and  possible numerical difficulties in the  determination of the collisional profiles.
The fact that the energetics of the blue wing of the line investigated
in the present
paper is  dominantly due to transitions between the  repulsive $\Sigma$ states (see below) somewhat legitimates the neglect of  spin-orbit  coupling for the line profile calculations, since these repulsive states are always  adiabatically correlated with the highest atomic asymptotes for a given $n$, with or without taking account of fine structure ($2p$ or $2p_{3/2}$ for n=2,
$3d$ or $3d_{5/2}$ for n=3). The evolution of the  dipole transition moment between the relevant $\Sigma$ repulsive states should thus be preserved.

\begin{figure}
 \centering
\resizebox{0.46\textwidth}{!}
{\includegraphics*{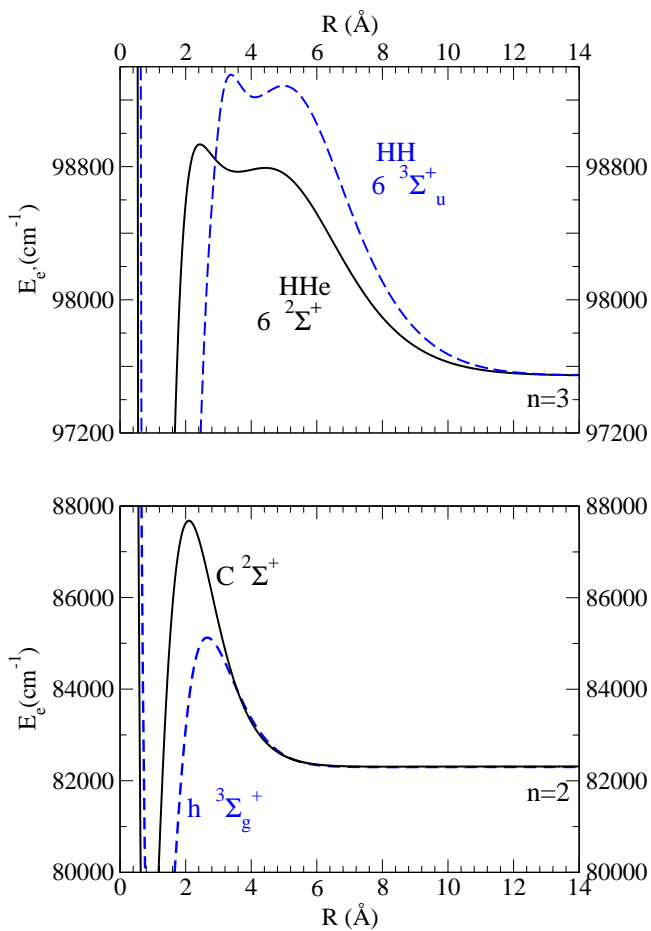}}
\caption  { Upper and lower repulsive potentials of H-He and H-H.
  Top: Short-range part of the repulsive potential curve 
  6~$^2\Sigma^{+}$ state (black line)
        of the H-He molecule compared  with the 
   6~$^3\Sigma_u^{+}$ state (blue dashed line)  of the H-H.
   Bottom:  Short-range part of the  potential curves of
        $C~^2\Sigma^{+}$ state (black line) of the H-He molecule compared  with the 
	$h$~$^3\Sigma_g^{+}$ state (blue dashed line)  of {H-H}.
   $E_{e}$ and $E_{e' }$ are  the potential energies  of  the initial ($n$=2)
   and final ($n$=3) atomic states of the transitions.}
\label{sec:pots}
\end{figure}

 \begin{figure}
 \centering
\resizebox{0.46\textwidth}{!}
{\includegraphics*{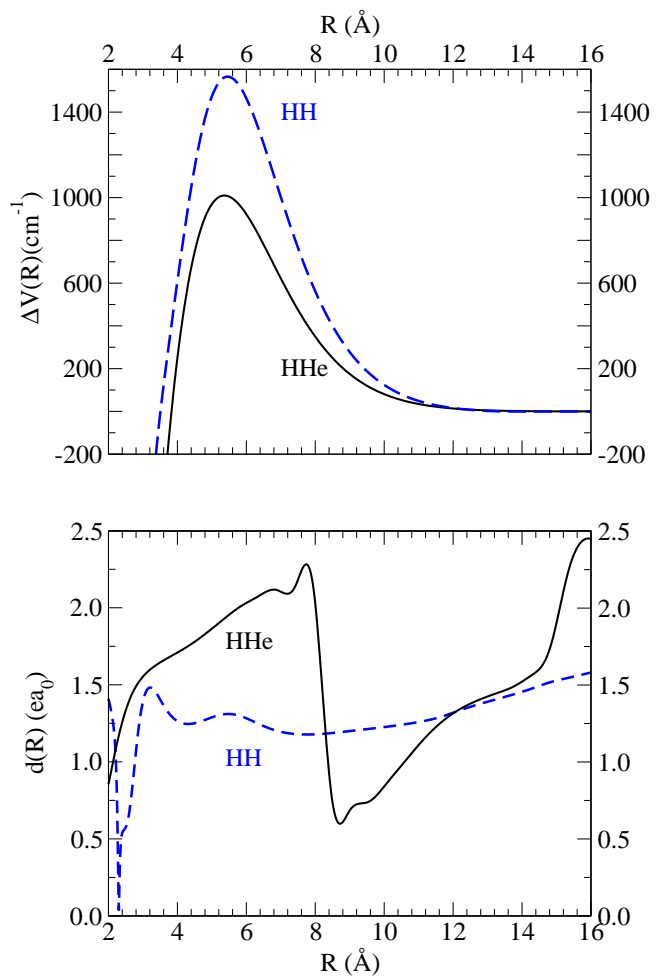}}
\caption  { Difference potentials and electric transition
    dipole moments related to the repulsive states of H-H and H-He.
  Top: Difference  potential for the triplet transition
\mbox{6~$^3\Sigma_u^{+}$-$h$~$^3\Sigma_g^{+}$} (blue dashed line) of H-H
         compared  with the doublet transition \mbox{6~$^2\Sigma^{+}$-$C~^2\Sigma^+$}  of
H-He (black line).
           Bottom: Electric dipole   moments of the triplet state transition  
  of H-H (blue dashed line) 
 and of the doublet state transition  of H-He (black line).}  
\label{sec:vdiff}
 \end{figure}

\section{Collisional profiles perturbed by  He atoms}
\label{sec:BaHHe}
The potentials and radiative dipole transition moments described above
are input data
for a unified spectral line shape evaluation of the Balmer line.
This treatment includes the finite duration of collision since
it is well known that the impact approximation, which assumes that collisions
occur instantaneously, causes the Lorentzian approximation to be inaccurate
far from the line center.
Although our unified theory was developed in \citet{allard1999},
 and a detailed discussion is presented there, we   provide  an overview
 in  Sect.~3 of our  preceding paper   on the
 Lyman-$\alpha$ line \citep{spiegelman2021}.
 In that paper we focussed     on the medium, long distance,
   and  asymptotic limit of the potential curves,  and we  were able to 
 solve the  uncertainty  remaining
in \citet{allard2020} involving the adiabatic asymptotic correlation of
the $\Sigma$ states in Lyman-$\alpha$.
 The bottom plot  of Fig.~\ref{fig:e3lr}   shows the long-range part of the
MRCI adiabatic H-He potential energy curves of  states  $A$, $B$, and $C$
dissociating into  H($n$=2)+ He.
As a result of the long-distance avoided crossing around 8.1~\AA\/  in
the $^2\Sigma^+$
manifold between the $A~^2\Sigma^+$ state and the $C~^2\Sigma^+$state, the latter is
adiabatically correlated to 2$p$.
This long-range avoided crossing was also shown to induce a  kink at 8.1~\AA\/
in dipole moment $X~^2\Sigma^+-C~^2\Sigma^+$ and a sign change in   the
$X~^2\Sigma^+-A~^2\Sigma^+$ dipole transition
moment (Fig.~3 in \citealt{spiegelman2021}).

\begin{figure}
\centering
\vspace{8mm}
\includegraphics[width=8cm]{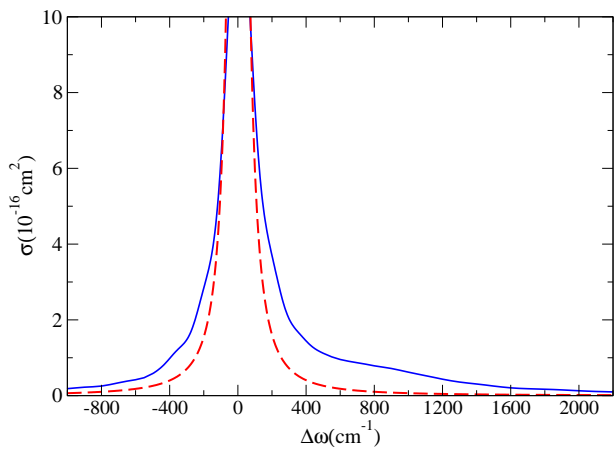}
\caption{Comparison  of the   Balmer-$\alpha$ line perturbed by
  H-He  collisions (blue curve)  with the Lorentzian approximation
  (red dashed  curve). The He density is 5$\times$10$^{20}$~cm$^{-3}$; 
  the  temperature is 8000~K. }
\label{BAHHeomg}
\end{figure}

\begin{figure}
\centering
\vspace{8mm}
\includegraphics[width=8cm]{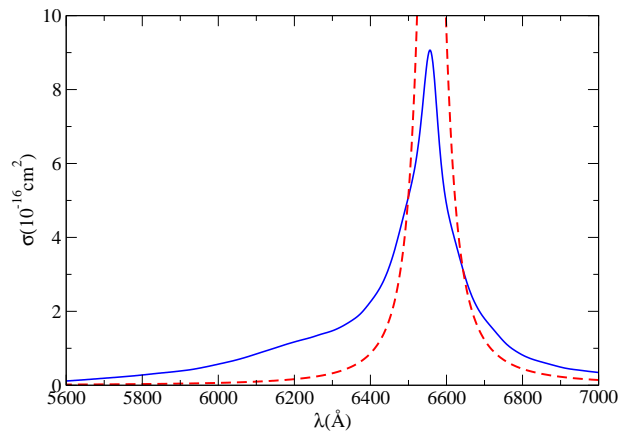}
\caption{Comparison  of the   Balmer-$\alpha$ line perturbed by
  H-He  collisions (blue curve)  with the   Lorentzian approximation
  (dashed red curve). The  He density is 10$^{21}$  cm$^{-3}$;
  the  temperature is 8000~K.}
\label{BAHHelam}
\end{figure}

Figure~\ref{sec:pots} shows the short-range part of the  potential
curve  $E_{e'}(R)$ of the  repulsive $3d$~6~$^2\Sigma^{+}$ state.
The prediction  of a line satellite in the blue wing of
the  H-He line profile is related to the potential maximum
of  $E_{e' }(R)$ in the 
internuclear distance range $R$=~2.5--5.5~\AA\/.
This leads to a maximum
of the potential energy difference $\Delta V(R)$ at 5.5~\AA\/
for the \hbox{ $C~^2\Sigma^+ \rightarrow 6~^2\Sigma^+$} transition.
 $\Delta V(R)$ shown in Fig.~\ref{sec:vdiff}   is given by:
\begin{equation}
\Delta V(R) \equiv V_{e' e}(R) = V_{e' }(R) - V_{ e}(R) \; ,
\label{eq:deltaV}
\end{equation}
and represents the difference  between the  energies
of the quasi-molecular transition.
The potential energy $V_{e}(R)$ for a state $e$ is defined as:
\begin{equation}
V_{e}(R) = E_e(R)-E_e^{\infty} \; .
\label{eq:V}
\end{equation}
The  unified theory  predicts that line satellites
 will be centered
 periodically at frequencies corresponding to integer multiples of
 the extrema of  $\Delta V(R)$.
 The satellite amplitude depends on the value of the electric dipole
 transition moment through the  region of the potential extremum responsible
 for the satellite and on the position of this extremum \citep{allard1998b}.
  The electric dipole transition moment is plotted in
 Fig.~\ref{fig:dipsig23} and  the bottom plot of Fig.~\ref{sec:vdiff}.
  The $C~^2\Sigma^+-6~^2\Sigma^+$ transition tends to the
  asymptotically  allowed transition $2p-3d$.
  In the present  study we  consider the temperature and density range
of cool DZA white dwarf stars.
Figure~\ref{BAHHeomg} shows
a very broad blue wing with a  close line  satellite
at about 900~cm$^{-1}$ from the line center corresponding to the
maximum of $\Delta V (R)$. The deviation from Lorentzian behavior is real, and
the asymmetry of the line profile becomes apparent not much farther from the line on the blue short-wavelength side.
There is a linear variation in the strength of the blue wing
with helium density as long as
 \mbox{ n$_{\rm He}$ $\leq$ 5$\times$ 10$^{20}$  cm$^{-3}$}; however, 
 when the He density is increased from 5$\times$10$^{20}$
to 10$^{21}$~cm$^{-3}$ the development of the blue wing 
 leads to the center of the main line being overwhelmed by the
 line satellite.
At 10$^{21}$~cm$^{-3}$ 
 the core of the line is shallow and no longer Lorentzian.
 Figure~\ref{BAHHelam}  clearly illustrates how 
 the non-Lorentzian shapes can exist because of the presence of close line
 satellites.
In these  physical conditions found in the cool atmosphere of
  DZA white dwarfs, we are 
 outside the  range of validity of the Lorentzian 
approximation usually used in stellar atmosphere modeling.
 This clear asymmetry in the wings of  the Balmer-$\alpha$ line was noticed by
Koester (1995, private communication) in the  low-resolution optical spectrum
of  the DZA white dwarf L745-46A obtained
 at the ESO La Silla 3.6m telescope.
Figure~3 in \citet{koester2000} shows the observed very shallow  Balmer-$\alpha$
line with a very broad blue wing.

\section{Collisional profiles perturbed by  H atoms}
\label{sec:BaHH}

We  used  the  non-relativistic  ab initio calculations of the  H-H
potentials of
Spielfiedel (2001, private communications), in which the asymptotic states
remain degenerate 
for a given hydrogen quantum number $n$.
Figures~1 and 2
in  \citet{allard2008a} show the potential
 energies correlated to the $3d$ and $2p$ states.  Essentially  20
transitions generate
the 3$d$-2$p$  H-H component that  gives the main contribution to
the Balmer-$\alpha$ line broadening.
Electronic singlet-singlet and triplet-triplet transition moments  of
H$_2$ have already been  very well studied  by
\citet{spielfiedel2003} and \citet{spielfiedel2004}.

\subsection{Self-broadening of the Balmer-$\alpha$ line}

In a unified treatment, the complete spectral 
  energy distribution is computed from the core to the far line wing.
The Lorentzian  widths and shifts can be readily extracted from our
unified line broadening calculations  \citep{allard1999}. 
 In \citet{allard2008a},
we  computed the width and shift of the Balmer-$\alpha$ line
perturbed by neutral hydrogen, and studied their dependence on a full range of 
 temperatures from 3000 to 12000~K needed for stellar spectra models.
Previous resonance broadening  parameters 
based on a multipole expansion of the interaction, neglecting  the 
van der Waals interactions,   had been calculated by  \citet{ali1966}.
Their work was widely adopted for use in stellar atmosphere
models. 
Our calculations lead to larger values than those
obtained with the commonly used theory of \citet{ali1966}
 and are closer to the  calculations of 
\citet{barklem2000a} and \citet{barklem2000b}.
Using up-to-date theories of the Balmer-$\alpha$ broadening mechanisms,
\citet{cayrel2011a} 
was  able to obtain excellent fits of observed  stellar Balmer-$\alpha$
profiles with 
computed profiles, using 1D Kurucz Atlas9 models, in the temperature range from 
5000K to  7000 K; the other parameters  (metallicity, $[\alpha/\mathrm{Fe}]$,
and log $g$) were fixed by a reliable detailed analysis of the atmosphere.
The new values of the collisional self-broadening of Balmer-$\alpha$ have 
raised a problem that had remained hidden because
of the far too small value of the cross-section proposed by Ali-Griem which 
 had been largely used since 1966. The accuracy of the new values excludes
 the former agreement between
 observed and computed Balmer-$\alpha$ profiles with 1D models.
  For Balmer series profiles in the Sun and generally in F, G, and K stars, only 3D radiative hydrodynamical models that rely on fundamental
   physics are adequate to model accurately the complexities of convection and the emergent line profile \citep{ludwig2009,amarsi2018,giribaldi2019}.
   Moreover, the shift and width of hydrogen Balmer-$\alpha$ at
   high electron density in a laser-produced plasma  reported
   in \citet{kielkopf2014} would suggest that for conditions of white dwarf
   stars  the atomic physics  is not completely  understood.

  \subsection{Blue wing  of the Balmer-$\alpha$ line}
  
\begin{figure}
\centering
\vspace{8mm}
\includegraphics[width=8cm]{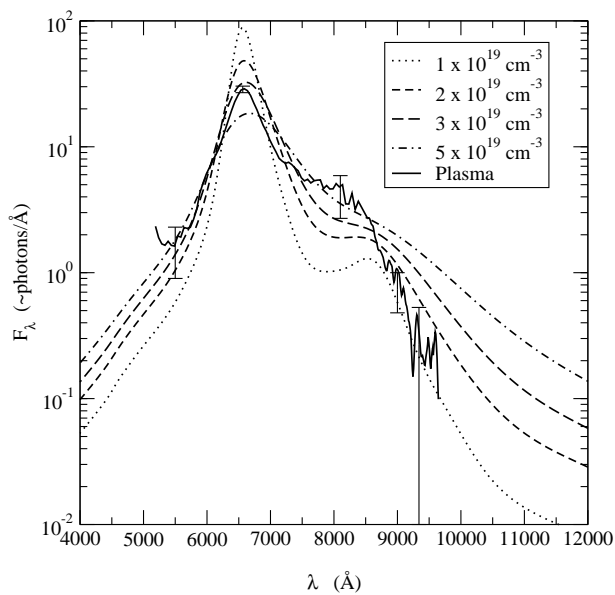} 
\caption{Comparison of the Balmer-$\alpha$ line as
seen in the time-integrated experimental data with theoretical profiles
at different ion densities  (extracted from  Fig.~8 of \citet{kielkopf2002}).
 The plasma density is 3$\times$ 10$^{19}$ cm$^{-3}$.}
\label{fig:ha893_comp}
\end{figure}

The unified theory of spectral line broadening applied to
  the Balmer-$\alpha$ line of atomic hydrogen predicts  structure in the 
Balmer-$\alpha$ line wing due to radiation that is emitted 
during atom-ion collisions. The 
strongest feature  is a satellite at 8650~\AA\/. Laboratory observations of
a laser-produced hydrogen plasma confirm that prediction \citep{kielkopf2002}.
In Fig.~\ref{fig:ha893_comp} we  compare the theoretical 
profiles in the region of the 8650~\AA\/  satellite due to H-H$^+$
interaction with an experimental profile. Line satellites can also be
predicted in the  Balmer-$\alpha$ line
perturbed by neutral hydrogen.

\begin{figure}
\centering
\vspace{8mm}
\includegraphics[width=8cm]{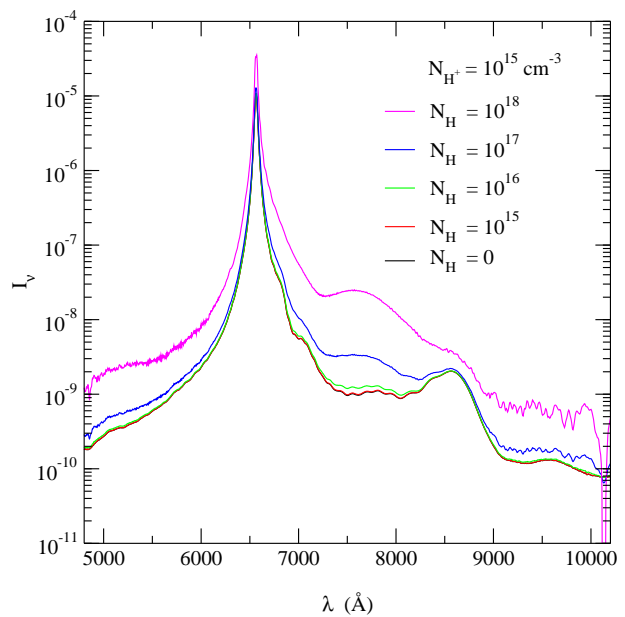}
\caption{Variation  of the   Balmer-$\alpha$ line perturbed by simultaneous 
   collisions by protons and neutral H atom.  
   The  H$^+$ density remains equal to  10$^{15}$  cm$^{-3}$ and the
   H density varies from 0 to  10$^{18}$  cm$^{-3}$. The temperature is 10000~K.}
\label{BAHHlam}
\end{figure}

\begin{figure}
\centering
\vspace{8mm}
\includegraphics[width=8cm]{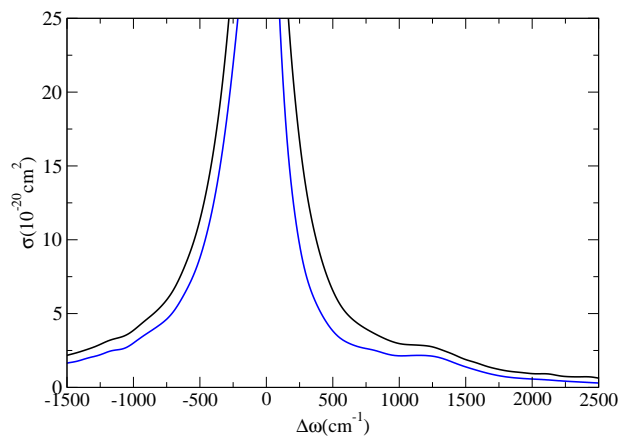}
\caption{ Comparison  of the   Balmer-$\alpha$ line perturbed by
    H-H  collisions (black curve) with  the contribution of  the triplet
    transitions alone  (blue curve). 
   The  H density is 10$^{18}$  cm$^{-3}$; the  temperature is 5000~K.}
\label{BAHHomg}
\end{figure}

\begin{figure}
\centering
\vspace{8mm}
\includegraphics[width=8cm]{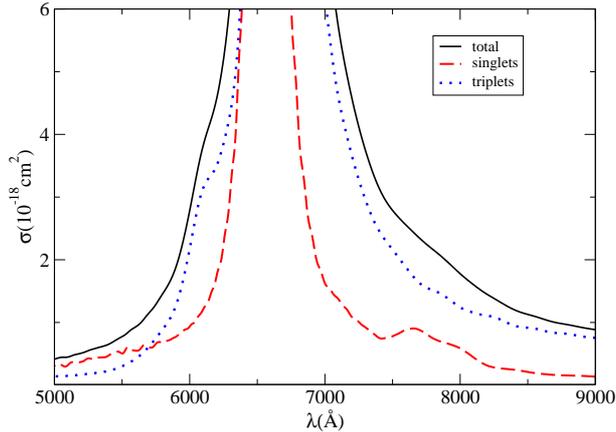}
\caption{ Balmer-$\alpha$ line, perturbed by H-H collisions (black curve),
    and the Balmer line with the sole contribution of the triplet
    transitions (blue dotted curve)
and the sole contribution of the singlet transitions (dashed red curve).
The   H density is 10$^{21}$  cm$^{-3}$; the  temperature is 5000~K.}
\label{BAHHlam21}
\end{figure}

The diatomic potentials and electronic transition moments among
singlets and triplets exist for the Balmer-$\alpha$ line
perturbed by collisions with neutral H; however,  we have not previously reported
our theoretical
study of the wings of Balmer-$\alpha$.
 Accounting for  all transitions between  $n_i=2$ and $n_f=3$ levels
  generates numerous molecular states 
that  lead to the same asymptotic energy difference at $R \rightarrow \infty$.
 Of the total number, 36 asymptotically allowed transitions
contribute to Balmer-$\alpha$ and 32 are forbidden.
For the Lyman-$\alpha$ line, the triplet transitions contribute
in the blue part of the profile and the singlet transitions in the red
part of the profile.
The line wings have been quantitatively examined for all of the components
that contribute to the Balmer-$\alpha$ line.
Our calculations allow us  to identify which
  transition leads to a line satellite.
Figure~\ref{BAHHlam} shows  the red satellite feature due to the
 \mbox{$C~1^1\Pi_u$ $\rightarrow 1^1\Delta_g$} centered at about 7800~\AA\/
in the red wing,  
whereas a close blue satellite is shown in Fig.~\ref{BAHHomg}.
We  restrict the present study to the  blue wing.
 Figure~\ref{sec:pots} shows the short-range part of the repulsive potential
curve of the  6~$^2\Sigma^{+}$ of the H-He molecule compared  with the 
6~$^3\Sigma_u^{+}$ state of the H$_2$ molecule. 
For H-H, the highest 6~$^3\Sigma_u^{+}$ is repulsive and leads to a maximum
in the difference  potential of the triplet transition
\mbox{$h$~$^3\Sigma_g^{+}$ $\rightarrow$ 6~$^3\Sigma_u^{+}$}
(Fig.~\ref{sec:vdiff}), which is responsible  for the formation of a blue line
satellite  at 1200~cm$^{-1}$.
The resulting asymmetry of the Balmer-$\alpha$ line can be clearly seen
in Fig.~\ref{BAHHomg}. The line shapes shown in  
Figs.~\ref{BAHHomg} and \ref{BAHHeomg} are
very similar, the blue H$_2$ line satellite being  a little farther
from the core of the line in the case of H-H collisions.
When the H density increases from 10$^{18}$  to 10$^{21}$~cm$^{-3}$
(Fig.~\ref{BAHHlam21}) the  satellite features
 appear as shoulders centered at about  7800~\AA\/ in the
red wing   and at 6100~\AA\/ in the blue wing. 
The H-H blue satellite is still apparent as a shoulder, whereas the H-He satellite is lost in the blue tail. 
They  both emphasize the non-Lorentzian behavior on the blue side.
The change from Lorentzian can be attributed entirely to radiation during close  H-H and H-He collisions.
In Fig.~11 in \citet{allard2008a} we compared our calculation of the 
unified theory line profile  to the Lorentzian profile
using the impact limit for the same  H density 10$^{18}$  cm$^{-3}$
as in  Fig.~\ref{BAHHomg}.
The Lorentzian profile shown in Fig.~11 of \citet{allard2008a}
is a useful representation of the unified line shape only
from 6540 to 6580~\AA\/.

\section{General trend  of the repulsive $\Sigma$ states and conclusion}
\label{sec:trend}

\citet{spiegelman2021} made an exhaustive study of the blue asymmetry
 in  the Lyman-$\alpha$  line profile perturbed by H-He and H-H collisions.
In the present paper we have shown that  a similar asymmetry  exists for the 
Balmer-$\alpha$ line profile.

\begin{figure}
\centering
\vspace{8mm}
\includegraphics[width=8cm]{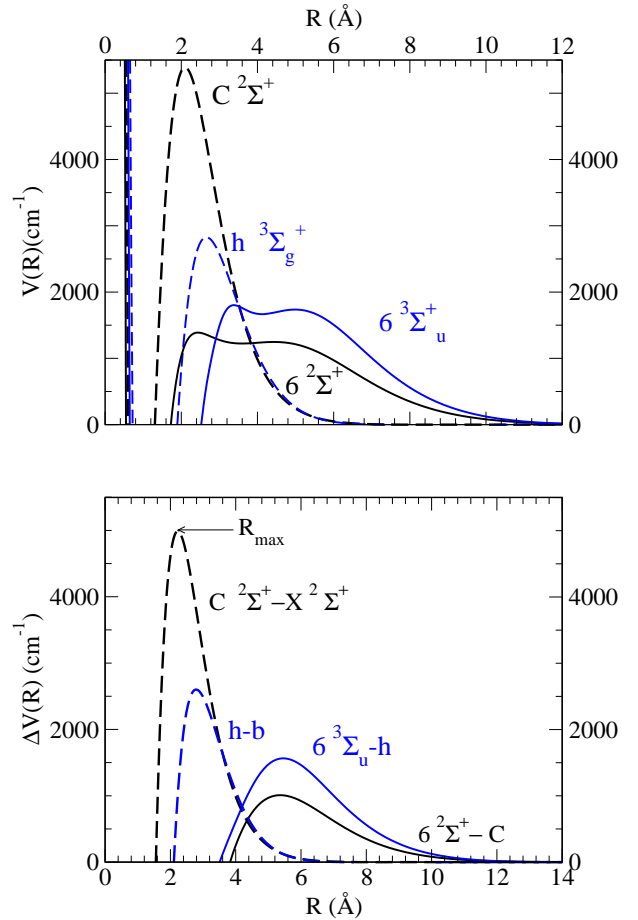}
\caption{ Repulsive potentials and their difference for H-He and H-H.
  Top: Potential energy $V$ of the repulsive
  states of Balmer-$\alpha$ perturbed by
  H-He  collisions (black curves); H-H  collisions (blue curves); and the  same, but for  Lyman-$\alpha$
  (dashed curves).
  Bottom: Difference  potential $\Delta V (R)$
  for the Balmer transitions 
6~$^2\Sigma^{+}$-$C~^2\Sigma^+$  of H-He collisions (black curves);
6~$^3\Sigma_u^{+}$-$h$~$^3\Sigma_g^{+}$ of H-H collisions (blue curves); and the same, but for the Lyman  (dashed curves)
$C~^2\Sigma^+$-$X~^2\Sigma^+$ and $h$~$^3\Sigma_g^{+}$-$b$~$^3\Sigma_u^{+}$
transitions.}
\label{sec:balym}
\end{figure}

Figure~\ref{sec:balym}  illustrates the  repulsive
potential energies and their difference $\Delta V (R)$ for the 
transitions $n_i$ $\rightarrow$  $n_f$ involved in Lyman-$\alpha$ and
Balmer-$\alpha$.  Their importance must be noted
in view of the study of the asymmetrical shape dependence on
the order of the Balmer series.
In the  case of the Lyman-$\alpha$ and Balmer-$\alpha$ lines,
 among all molecular states correlated to a given atomic configuration,
 the highest ones in the configuration,
namely \mbox{ $C~^2\Sigma^+$ and 6~$^2\Sigma^+$}  for H-He and
\mbox{ h~$^3\Sigma_g^{+}$ and 6~$^3\Sigma_u^{+}$} for H-H,
are repulsive and have a prominent role in the appearance of features  in
the blue wing.

 For \mbox{ $n_i$=2 $\rightarrow$  $n_f$} with $n_f$  $>$ 3, the transitions that  give rise to a blue satellite will be asymptotically forbidden.

Due to the more diffuse character of the 3$s$, 3$p$, and 3$d$ orbitals
in comparison with the 2$s$ and  2$p$, the barrier of state
6~$^2\Sigma^+$ extends
to a wider  distance range  and is lower than that of state
$C~^2\Sigma^+$ (Fig.~\ref{sec:balym}, top).
 The bottom plot of Fig.~\ref{sec:balym} shows that 
the maximum in $\Delta V$  occurs at larger
 internuclear distances ($R_{\rm max}$ $\sim$ 5.4~\AA\/) for Balmer-$\alpha$ than for Lyman-$\alpha$.
This leads to an  increase in the collision volume;
the average number of perturbers in the
interaction volume at $R_{\rm max}$ is the determining parameter for
the amplitude of
the satellites on the spectral line \citep{allard1978,royer1978,allard1982}.
This dependence on the average number of
perturbers in the collision volume is expected on the basis
of the Poisson distribution, which indicates the probability
of finding a given  number of uncorrelated perturbers in
the collision volume.
It was identified  decisively in the theoretical analysis of experimental Cs spectra 
by \citet{kielkopf1979}.

For  the H-He Balmer-$\alpha$ line, the maximum $\Delta V_{\rm{}max}$ is 
 smaller, 1000~cm$^{-1}$
versus 5000 cm$^{-1}$ for the Lyman-$\alpha$ line (Fig.~\ref{sec:balym}, bottom). 
The Balmer line satellite is then closer  to the main line than 
the Lyman satellite, as shown in 
Fig.~\ref{BAHHeomg} and Fig.~11 of \citet{spiegelman2021}.
Because of this general trend  characterizing the repulsive $\Sigma$ states,
 the average number of perturbers in the interaction 
 volume will be  larger for higher series of Balmer lines
  leading to a higher probability of multiple pertuber
  effects that give rise to the long blue tail due to the
 closer relative proximity of the satellite band to the line.

   This effect can be unimportant when the perturber density 
    is low, whereas it becomes dramatic  when this density can be as large as
    10$^{21}$  cm$^{-3}$ in the cool atmosphere of
    DZA white dwarfs.
   The Balmer-$\alpha$ line profile shown in Fig.~\ref{BAHHlam21} is
    purely  formal as we do not reach this H density in cool stars,
    but was essentially done for our comparison H-He versus H-H.

For the higher series of Balmer lines we can predict that the main 
line is not a Lorentzian, but  is replaced by a blend of
multiple satellites~\citep{kielkopf1983,kielkopf1985}.
We have shown (Fig.~\ref{BAHHelam}) that the  Balmer-$\alpha$ line
core is  no longer a Lorentzian when the  He density is as high
as 10$^{21}$~cm$^{-3}$. 

When the helium density is as large as it is in the DZA stars such as
L745-46A, only  the Balmer-$\alpha$ line is observed and it is very shallow.
For cooler WD  when the helium density is even higher, Balmer lines are
extremely weak or totally absent as in extreme helium stars.
We have conclusively shown the need for precise accurate and complete profiles of neutral collision-broadened lines when hydrogen abundances are found from
their profiles in stellar spectra.

\begin{acknowledgements}
 NFA is grateful to M. Leseignoux, C. Giovanetti, A. Lekic, F. Lanzi,
  B. Pape and A. Dereu for their help
  in the analysis of  the huge number of molecular transitions
  which contribute to the Balmer-$\alpha$ line
  perturbed by collisions with neutral H.
   We thank the referee for helpful comments and P. Bonifacio for references
  of 3D calculations.
\end{acknowledgements}


\begin{thebibliography}{46}
\expandafter\ifx\csname natexlab\endcsname\relax\def\natexlab#1{#1}\fi

\bibitem[{{Ali} \& {Griem}(1966)}]{ali1966}
{Ali}, A.~W. \& {Griem}, H.~R. 1966, Physical Review, 144, 366

\bibitem[{Allard(1978)}]{allard1978}
Allard, N.~F. 1978, J. Phys. B: At. Mol. Opt. Phys., 11, 1383

\bibitem[{Allard {et~al.}(1998)Allard, Drira, Gerbaldi, Kielkopf, \&
  Spielfiedel}]{allard1998b}
Allard, N.~F., Drira, I., Gerbaldi, M., Kielkopf, J.~F., \& Spielfiedel, A.
  1998, A\&A, 335, 1124

\bibitem[{Allard \& Kielkopf(1982)}]{allard1982}
Allard, N.~F. \& Kielkopf, J.~F. 1982, Rev. Mod. Phys., 54, 1103

\bibitem[{Allard {et~al.}(2007)Allard, Kielkopf, \& Allard}]{allard2007c}
Allard, N.~F., Kielkopf, J.~F., \& Allard, F. 2007, EPJ D, 44, 507

\bibitem[{Allard {et~al.}(2008)Allard, Kielkopf, Cayrel, \& van~'t
  Veer-Menneret}]{allard2008a}
Allard, N.~F., Kielkopf, J.~F., Cayrel, R., \& van~'t Veer-Menneret, C. 2008,
  A\&A, 480, 581

\bibitem[{{Allard} {et~al.}(2020){Allard}, {Kielkopf}, {Xu}, {Guillon},
  {Mehnen}, {Linguerri}, {Al Mogren}, {Hochlaf}, \& {Hubeny}}]{allard2020}
{Allard}, N.~F., {Kielkopf}, J.~F., {Xu}, S., {et~al.} 2020, \mnras, 494, 868

\bibitem[{Allard {et~al.}(1999)Allard, Royer, Kielkopf, \&
  Feautrier}]{allard1999}
Allard, N.~F., Royer, A., Kielkopf, J.~F., \& Feautrier, N. 1999, Phys. Rev. A,
  60, 1021

\bibitem[{{Amarsi} {et~al.}(2018){Amarsi}, {Nordlander}, {Barklem}, {Asplund},
  {Collet}, \& {Lind}}]{amarsi2018}
{Amarsi}, A.~M., {Nordlander}, T., {Barklem}, P.~S., {et~al.} 2018, \aap, 615,
  A139

\bibitem[{{Barklem} {et~al.}(2000{\natexlab{a}}){Barklem}, {Piskunov}, \&
  {O'Mara}}]{barklem2000b}
{Barklem}, P.~S., {Piskunov}, N., \& {O'Mara}, B.~J. 2000{\natexlab{a}}, \aap,
  363, 1091

\bibitem[{{Barklem} {et~al.}(2000{\natexlab{b}}){Barklem}, {Piskunov}, \&
  {O'Mara}}]{barklem2000a}
{Barklem}, P.~S., {Piskunov}, N., \& {O'Mara}, B.~J. 2000{\natexlab{b}}, \aap,
  355, L5

\bibitem[{{Barklem} {et~al.}(2002){Barklem}, {Stempels}, {Allende Prieto},
  {Kochukhov}, {Piskunov}, \& {O'Mara}}]{barklem2002}
{Barklem}, P.~S., {Stempels}, H.~C., {Allende Prieto}, C., {et~al.} 2002, \aap,
  385, 951

\bibitem[{Brooks \& Hunt(1988)}]{brooks88}
Brooks, R.~L. \& Hunt, J.~L. 1988, The Journal of Chemical Physics, 89, 7077

\bibitem[{{Cayrel} {et~al.}(2011){Cayrel}, {van't Veer-Menneret}, {Allard}, \&
  {Stehl{\'e}}}]{cayrel2011a}
{Cayrel}, R., {van't Veer-Menneret}, C., {Allard}, N.~F., \& {Stehl{\'e}}, C.
  2011, \aap, 531, A83

\bibitem[{{Cukanovaite} {et~al.}(2021){Cukanovaite}, {Tremblay}, {Bergeron},
  {Freytag}, {Ludwig}, \& {Steffen}}]{cukanovaite2021}
{Cukanovaite}, E., {Tremblay}, P.-E., {Bergeron}, P., {et~al.} 2021, \mnras,
  501, 5274

\bibitem[{{Giribaldi} {et~al.}(2019){Giribaldi}, {Ubaldo-Melo}, {Porto de
  Mello}, {Pasquini}, {Ludwig}, {Ulmer-Moll}, \&
  {Lorenzo-Oliveira}}]{giribaldi2019}
{Giribaldi}, R.~E., {Ubaldo-Melo}, M.~L., {Porto de Mello}, G.~F., {et~al.}
  2019, \aap, 624, A10

\bibitem[{Ketterle(1989)}]{ketterle89}
Ketterle, W. 1989, Phys. Rev. Lett., 62, 1480

\bibitem[{Ketterle(1990{\natexlab{a}})}]{ketterle90a}
Ketterle, W. 1990{\natexlab{a}}, The Journal of Chemical Physics, 93, 3752

\bibitem[{Ketterle(1990{\natexlab{b}})}]{ketterle90b}
Ketterle, W. 1990{\natexlab{b}}, The Journal of Chemical Physics, 93, 3760

\bibitem[{Ketterle(1990{\natexlab{c}})}]{ketterle90d}
Ketterle, W. 1990{\natexlab{c}}, The Journal of Chemical Physics, 93, 6935

\bibitem[{Ketterle {et~al.}(1988)Ketterle, Dodhy, \& Walther}]{ketterle88}
Ketterle, W., Dodhy, A., \& Walther, H. 1988, The Journal of Chemical Physics,
  89, 3442

\bibitem[{Ketterle {et~al.}(1985)Ketterle, Figger, \& Walther}]{ketterle85}
Ketterle, W., Figger, H., \& Walther, H. 1985, Phys. Rev. Lett., 55, 2941

\bibitem[{{Kielkopf}(1983)}]{kielkopf1983}
{Kielkopf}, J. 1983, Journal of Physics B Atomic Molecular Physics, 16, 3149

\bibitem[{{Kielkopf}(1985)}]{kielkopf1985}
{Kielkopf}, J. 1985, \jqsrt, 33, 267

\bibitem[{{Kielkopf} \& {Allard}(1979)}]{kielkopf1979}
{Kielkopf}, J.~F. \& {Allard}, N.~F. 1979, Physical Review Letters, 43, 196

\bibitem[{{Kielkopf} \& {Allard}(2014)}]{kielkopf2014}
{Kielkopf}, J.~F. \& {Allard}, N.~F. 2014, Journal of Physics B Atomic
  Molecular Physics, 47, 155701

\bibitem[{{Kielkopf} {et~al.}(2002){Kielkopf}, {Allard}, \&
  {Decrette}}]{kielkopf2002}
{Kielkopf}, J.~F., {Allard}, N.~F., \& {Decrette}, A. 2002, European Physical
  Journal D, 18, 51

\bibitem[{Knowles \& Werner(1992)}]{knowles92}
Knowles, P. \& Werner, H.-J. 1992, Theoretica Chim. Acta, 84, 95–103

\bibitem[{{Koester} \& {Wolff}(2000)}]{koester2000}
{Koester}, D. \& {Wolff}, B. 2000, \aap, 357, 587

\bibitem[{Kramida {et~al.}(2020)Kramida, {Yu.~Ralchenko}, Reader, \& {and NIST
  ASD Team}}]{nist2020}
Kramida, A., {Yu.~Ralchenko}, Reader, J., \& {and NIST ASD Team}. 2020, {NIST
  Atomic Spectra Database (ver. 5.8), [Online]. Available:
  {\tt{https://physics.nist.gov/asd}} [2021, January 26]. National Institute of
  Standards and Technology, Gaithersburg, MD.}

\bibitem[{Lo {et~al.}(2006)Lo, Klobukowski, Bieli{\'{n}}ska-Waz, Schreiner, \&
  Diercksen}]{lo2006}
Lo, J. M.~H., Klobukowski, M., Bieli{\'{n}}ska-Waz, D., Schreiner, E. W.~S., \&
  Diercksen, G. H.~F. 2006, Journal of Physics B: Atomic, Molecular and Optical
  Physics, 39, 2385

\bibitem[{{Ludwig} {et~al.}(2009){Ludwig}, {Behara}, {Steffen}, \&
  {Bonifacio}}]{ludwig2009}
{Ludwig}, H.~G., {Behara}, N.~T., {Steffen}, M., \& {Bonifacio}, P. 2009, \aap,
  502, L1

\bibitem[{Nakajima \& Hirao(2011)}]{nakajima2011}
Nakajima, T. \& Hirao, K. 2011, Chemical reviews, 112, 385

\bibitem[{{Peach}(2011)}]{peach2011}
{Peach}, G. 2011, Baltic Astronomy, 20, 516

\bibitem[{Petsalakis {et~al.}(1990)Petsalakis, Theodorakopoulos, \&
  Buenker}]{petsalakis92}
Petsalakis, I.~D., Theodorakopoulos, G., \& Buenker, R.~J. 1990, The Journal of
  Chemical Physics, 92, 4920

\bibitem[{Reiher(2006)}]{reiher2006}
Reiher, M. 2006, Theoretical Chemistry Accounts, 116, 241–252

\bibitem[{Royer(1978)}]{royer1978}
Royer, A. 1978, Acta Phys. Pol. A, 54, 805

\bibitem[{Sarpal {et~al.}(1991)Sarpal, Branchett, Tennyson, \&
  Morgan}]{sarpal1991}
Sarpal, B.~K., Branchett, S.~E., Tennyson, J., \& Morgan, L.~A. 1991, Journal
  of Physics B: Atomic, Molecular and Optical Physics, 24, 3685

\bibitem[{Spiegelman {et~al.}(2021)Spiegelman, Allard, \&
  Kielkopf}]{spiegelman2021}
Spiegelman, F., Allard, N., \& Kielkopf, J. 2021, A\&A, 651, A51

\bibitem[{Spielfiedel(2003)}]{spielfiedel2003}
Spielfiedel, A. 2003, J. Mol. Spectrosc., 217, 162

\bibitem[{Spielfiedel {et~al.}(2004)Spielfiedel, Palmieri, \&
  Mitrushenkov}]{spielfiedel2004}
Spielfiedel, A., Palmieri, P., \& Mitrushenkov, A. 2004, Molec. Phys., 102,
  2249

\bibitem[{Theodorakopoulos {et~al.}(1987)Theodorakopoulos, Petsalakis,
  Nicolaides, \& R.J.Buenker}]{theodora1987}
Theodorakopoulos, G., Petsalakis, I.~D., Nicolaides, C.~A., \& R.J.Buenker.
  1987, J. Phys. B, 20, 2339

\bibitem[{{Unsold}(1955)}]{unsold1955}
{Unsold}, A. 1955, {{Physik der Sternatmosph\"{a}ren MIT besonderer
  Ber\"{u}cksichtigung der Sonne}} (Berlin, Springer, 1955.~2.~Aufl.)

\bibitem[{van Hemert \& Peyerimhoff(1991)}]{vanhemert91}
van Hemert, M.~C. \& Peyerimhoff, S.~D. 1991, The Journal of Chemical Physics,
  94, 4369

\bibitem[{Werner {et~al.}(2015)Werner, Knowles, Knizia, Manby, {Sch\"{u}tz},
  {et~al.}}]{molpro2015}
Werner, H.-J., Knowles, P.~J., Knizia, G., {et~al.} 2015, MOLPRO, version
  2015.1, a package of ab initio programs

\bibitem[{{Xu} {et~al.}(2017){Xu}, {Zuckerman}, {Dufour}, {Young}, {Klein}, \&
  {Jura}}]{xu2017}
{Xu}, S., {Zuckerman}, B., {Dufour}, P., {et~al.} 2017, \apjl, 836, L7

\end{thebibliography}

\begin{appendix}

\onecolumn

\section{Spectroscopic constants of molecular states of HeH correlated
    with H(n=3)}
  
\vspace{0.5cm}
\begin{center}
                    {      Table A.1. Spectroscopic constants of  HeH
          adiabatic molecular states dissociating into  H(3s,3p,3d)+He(1s$^2$).}
          
\vspace*{0.3cm}          
\begin{tabular}{ccccccccc}
\hline
State & ref &  $R_e$ (\AA) & $\omega_e$ (cm$^{-1}$) & $\omega_{exe}$ (cm$^{-1}$) & $\omega_{eye}$ (cm$^{-1}$) & $D_e$ (eV/cm $^ {-1}$) & $T_e$ (cm$^{-1}$) & $T_{00}$ (cm$^{-1}$)\\
        \hline
                && &\\
                $4^2\Sigma^+$& (a) & 0.7640 & 3382.7  &162.8& 42.4& 2.172/17517.8&18382.2 & 18207.0   \\
                &(b)& 0.7675& 3383 && & & \\
                &(c)& 0.7677& 3187 && & & \\
                &(d)& 0.7702& 3405& 201&&2.16\\
                             & (e) & 0.7709 & 3346.0  &167.3&     &  2.139/17255.6\\
                &(f)         &     &        &         &&&                     &18215.24\\
                && &\\
                $5^2\Sigma^+$& (a)& 0.7759 & 3228.1 & 148.4 & 57.5 &2.066/16674.1& 19225.6 &  18979.3  \\
                &(c)&0.7775& 3057 & \\
                &(d)& 0.7816& 3252& 208&& 2.06\\
                             & (e) & 0.7832 & 3148.5  &141.1&      & 2.032/16394.8\\
                &(f)         &     &        &         &&&                     &18945.92\\
                && &\\
                $6^2\Sigma^+$ & (a)&  0.7851 & 3148.6 & 168.3& 31.1  & 1.680/13554.5 & 22345.7& 22050.7 \\
                             & (e) & 0.7921 & 3094.3  &149.5&     &  1.6522/13321.7\\
                && &\\
                $2^2\Pi$& (a)& 0.7727 & 3267.6 & 178.4&  22.1 & 2.096/16908& 18991.9 &    18752.7 \\
                &(b)&0.7788& 3233 & \\
                &(c)&0.7755& 3083 & \\
                &(d)& 0.7783& 3299& 207&& 2.08\\
                             & (e) & 0.7833 & 3157.3  &152.7&     &  2.018/16274.7\\
                &(f)         &     &        &    &&     &                     &18762.32\\
                && &\\
                $3^2\Pi$& (a)& 0.7750 & 3239.2 & 175.1&  24.6  & 2.057/ 16588 &19311.9 &  19059.7\\
                             & (e) & 0.7799 & 3174.2  &143.3&    &   2.045/16496.1\\
                &(f)         &     &        &     &&    &                     &19023.31\\
                && &\\
                $1^2\Delta$& (a)& 0.7738 & 3251.2 &177.4 & 23.3  & 2.034/16406 &19506.0&     19259.0 \\
                             & (e) & 0.7810 & 3195.1  &154.3&    &   1.986/16020.8\\
                &(f)         &     &        &     &&    &                     &19248.90\\
                && &\\
        \hline
\end{tabular}
\end{center}
        {\footnotesize Electronic dissociation energies $D_e$ are  calculated  with reference to  the respective  dissociation limits of the adiabatic states.  $T_e$ is  the electronic transition energy from state $A^2\Sigma^+$, $T_{00}$ the associated $v'$=0 to $v"$=0 transition. The present vibrational data were determined for $^4$HeH through  five-point polynomial interpolation. References: (a) theory, this work; (b) theory, \citet{theodora1987}; (c) theory, \citet{sarpal1991}; (d) theory, \citet{lo2006}; (e) theory, \citet{vanhemert91}; (f) experiment, \citet{ketterle90b}.}

\end{appendix}

\end{document}